\documentstyle[prl,aps]{revtex}

\newcommand{\bea}{\begin{eqnarray}}
\newcommand{\eea}{\end{eqnarray}}
\newcommand{\beq}{\begin{equation}}
\newcommand{\eeq}{\end{equation}}
\newcommand{\del}{\partial}
\newcommand{\lishi}{\langle\!\langle}
\newcommand{\rishi}{\rangle\!\rangle}

\begin{document}

\draft

\twocolumn[\hsize\textwidth\columnwidth\hsize\csname
@twocolumnfalse\endcsname



\title{Modular transformation and boundary states in 
logarithmic conformal field theory}

\author{Shinsuke Kawai
\footnotemark[1] 
and John F. Wheater
\footnotemark[2]
}
\address{Theoretical Physics, Department of Physics, 
 University of Oxford, 1 Keble Road, Oxford OX1 3NP, UK}
\date{\today}
\maketitle

\begin{abstract}

We study the $c=-2$ model of logarithmic conformal field theory in the 
presence of a boundary using symplectic fermions.
We find boundary states with consistent modular properties.
A peculiar feature of this model is that the vacuum representation
corresponding to the identity operator is a sub-representation of a ``reducible
but indecomposable'' larger representation.
This leads to unusual properties, such as the failure of the Verlinde formula.
Despite such complexities in the structure of modules,
our results suggest that logarithmic conformal field theories admit bona 
fide boundary states.
\end{abstract}

\noindent
\preprint{OUTP***, KUCP***, hep-th/0103197
}

\vskip2pc]
\footnotetext[1]{\tt s.kawai1@physics.ox.ac.uk}
\footnotetext[2]{\tt j.wheater1@physics.ox.ac.uk}

\section{Introduction}

The existence of critical systems with logarithmic correlation functions
has been known for some time.
It was almost a decade ago when Saleur\cite{saleur} pointed out the existence
of logarithmic behavior in the correlation functions of some geometrical 
critical systems. 
Subsequently, the importance of such logarithmic terms in conformal field 
theories (CFTs)
was emphasized by Gurarie\cite{gurarie}, who pointed out that this phenomenon
occurs when two (or more) primary operators have scaling dimensions
differing by integers, and that these operators are not diagonal under
conformal transformations but they form Jordan cells. 
Such theories had never been studied seriously till then, mainly because they
are not unitary. 
Importantly, however, non-unitary does not mean unrealistic -- there are a 
large number of non-unitary models relevant in statistical and 
high-energy physics. 
Such logarithmic conformal field theories (LCFTs) are now known to be quite 
ubiquitous\cite{cardy99}, and are being studied not only for their 
mathematical interests, 
but also for their potential applicability in string theory and statistical 
physics.

Although much work has been done in recent years, our knowledge of LCFTs is 
still incomplete.
In particular, the behavior of such systems in the presence of a boundary is 
not fully understood. In ordinary CFTs, the system near the 
boundary is described by bulk and boundary operators. 
Then, once the physical boundary states are identified, the scaling properties 
of the given system are completely understood. 
At least in the case of unitary minimal models, such physically consistent 
boundary states are most systematically obtained by the ``fusion 
method'' of Cardy\cite{cardy89}. 
This method uses the duality of open-string and closed-string channels; 
one partition function is calculated in two different ways, and consistency
gives non-trivial constraints on the boundary states which are realized 
physically. 
In LCFTs, however, the validity of this method, and even the existence of 
well-defined physical boundary states, has not been clear. 
The main obstacle is that operators in LCFTs are not classified into simple 
conformal towers which split the partition function into Virasoro
characters\cite{KoganWheater}. 
The purpose of this note is to show that Cardy's method {\em is} applicable
to LCFTs and that LCFTs allow physically consistent boundary states.

In this paper we focus on the triplet model of $c=-2$ which has explicit 
Lagrangian representation\cite{saleur,curious}:
\beq
S=\frac{1}{\pi}\int d^2z (\eta\bar\partial\xi+\bar\eta\partial\bar\xi),
\label{eqn:action}
\eeq
where $\eta$ and $\xi$ are fermionic fields of scaling dimensions $1$ and 
$0$, respectively, with operator products
$\eta(z)\xi(w)\sim\xi(z)\eta(w)\sim1/(z-w)$. 
This is nothing but the fermionic ghost system with central charge $-2$.
Our strategy is to derive explicit boundary states in Fock space and then to
study their modular properties.
Before discussing boundary states, we shall briefly summarize how the bulk
chiral theory is constructed from the action 
(\ref{eqn:action}) and thereby define our model precisely\cite{curious,sf}.
The energy-momentum tensor of this system is $T(z)=:\del\xi(z)\eta(z):$, 
from which the central charge is read off as $c=-2$. 
The system has a $U(1)$ symmetry whose current is 
$J_{U(1)}(z)= :\xi(z)\eta(z):$.
Due to this symmetry the system has twisted, as well as untwisted, sectors.
In this paper we consider the $Z_2$ orbifold model, where the twist 
operation is restricted to $Z_2$, which is a finite Abelian subgroup of $U(1)$.
The two fields $\eta$ and $\xi$ are conveniently combined into a two-component 
symplectic fermion\cite{curious,sf} of conformal dimension $1$ as 
\beq
\chi^+=\eta,\;\;\chi^-=\del\xi.
\eeq
This $Z_2$ orbifold symplectic fermion differs from the $\eta$-$\xi$ system
(\ref{eqn:action}) in the treatment of the $\xi$ zero mode, and its bosonic 
sector has been shown to be identical to the so-called triplet model\cite{sf}, 
which is one of the few models of LCFT that have been studied extensively.
Mode operators are defined by
$\chi^\pm(z)
=\sum_{m\in Z+\lambda}\chi^\pm_m z^{-m-1}$
and
$\bar\chi^\pm(\bar z)
=\sum_{m\in Z+\lambda}\bar\chi^\pm_m\bar z^{-m-1}$,
where $\lambda=0$ for the untwisted (Neveu-Schwartz) sector and $\lambda=1/2$ 
for the $Z_2$-twisted (Ramond) sector. 
Anticommutation relations consistent with the operator product of $\eta$ 
and $\xi$ are 
\beq
\{\chi^\alpha_m,\chi^\beta_n\}=md^{\alpha\beta}\delta_{m+n},
\label{eqn:sfcom}
\eeq 
where $d^{\alpha\beta}$ is an antisymmetric tensor such that 
$d^{\pm\mp}=\pm 1$, and $d_{\alpha\beta}$ is defined by 
$d^{\alpha\gamma}d_{\gamma\beta}=\delta^\alpha_\beta$.
The normal ordering in the twisted sector is defined on the twisted 
vacua, and the formulae are obtained using the operator products of the 
symplectic fermions with twist operators\cite{saleur,sf,dfms}, or resorting 
to the twisted Borcherds identity\cite{kac,ghost} in operator formalism. 
Then the energy-momentum tensor in terms of the generalized normal ordering 
becomes
$
T(z)= d_{\alpha\beta}:\chi^\alpha(z)\chi^\beta(z):/2
+\lambda(\lambda-1)/2,
$
and from this, the Virasoro operators are given as 
\beq
L_n=\frac 12 d_{\alpha\beta}\sum_{m\in Z+\lambda}
:\chi^\alpha_m\chi^\beta_{n-m}:
+\frac{\lambda(\lambda-1)}{2}\delta_{n0}.
\label{eqn:virop}
\eeq

In the twisted sector, the ground state is unique and is denoted by 
$\mu$. 
This has weight $-1/8$ and is identified as the ground state of the 
representation ${\cal V}_{-1/8}$. 
There is another representation ${\cal V}_{3/8}$ in this sector, 
which is built on the state $\nu^\alpha=\chi^\alpha_{-1/2}\mu$,
with weight $3/8$. 
The untwisted sector is more complicated because the existence of a zero-mode
leads to quadruply degenerate ground states. 
Let $\omega$ be a state annihilated by $\chi^\alpha_{m>0}$. 
Then, the four degenerate ground states are
$\omega$, $\theta^\pm=-\chi^\pm_0\omega$ and 
$\Omega=\chi^-_0\chi^+_0\omega=L_0\omega$.
Note that $\Omega$ is annihilated by further operations of zero-modes, and
hence identified as the M\"obius invariant vacuum. 
The bosonic ground states $\Omega$ and $\omega$ span a two-dimensional Jordan
cell on the action of $L_0$, 
and they are said to form a logarithmic pair.
Then $\Omega$ and $\omega$ constitute an irreducible vacuum 
representation ${\cal V}_0$, which is a sub-representation of the ``reducible 
but indecomposable\cite{rational,rohsiepe}'' representation ${\cal R}_0$.
There is another such logarithmic pair, $\phi^\alpha=\chi^\alpha_{-1}\omega$ 
and $\psi^\alpha=\chi^\alpha_{-1}\Omega$, constituting an irreducible 
representation ${\cal V}_1$, which is again included in a reducible but 
indecomposable larger representation ${\cal R}_1$. 
These six representations, ${\cal V}_{-1/8}$, ${\cal V}_{3/8}$, ${\cal V}_{0}$,
${\cal V}_{1}$, ${\cal R}_{0}$ and ${\cal R}_{1}$, are the building blocks of 
the $Z_2$ orbifold $c=-2$ model. 
We emphasize that these six are not all independent since 
${\cal V}_0$ and ${\cal V}_1$ are sub-representations of 
${\cal R}_0$ and ${\cal R}_1$, respectively. 
The characters for these representations are calculated in 
\cite{curious,sf,flohr} as:
\bea
&&\chi_{{\cal R}_0}(\tau)=\chi_{{\cal R}_1}(\tau)=2\Lambda_{1,2}(\tau),
\nonumber\\
&&\chi_{{\cal V}_0}(\tau)
=\frac 12\Lambda_{1,2}(\tau)+\frac 12 \eta(\tau)^2,
\nonumber\\
&&\chi_{{\cal V}_1}(\tau)
=\frac 12\Lambda_{1,2}(\tau)-\frac 12 \eta(\tau)^2,
\nonumber\\
&&\chi_{{\cal V}_{-1/8}}(\tau)=\Lambda_{0,2}(\tau),
\nonumber\\
&&\chi_{{\cal V}_{3/8}}(\tau)=\Lambda_{2,2}(\tau),
\label{eqn:char}
\eea
where
$\Lambda_{mn}(\tau)=\Theta_{mn}(\tau)/\eta(\tau)$.
We define the modular parameter as $q=e^{2\pi i\tau}$, and denote
the Jacobi $\Theta$-function and Dedekind $\eta$-function by
$\Theta_{mn}(\tau)=\sum_{k\in Z}q^{(2nk+m)^2/4n}$ and
$\eta(\tau)=q^{1/24}\prod_{k\in Z_+}(1-q^k)$, respectively.
The characters (\ref{eqn:char}) do not close under the modular 
transformation $\tau\rightarrow\tilde\tau=-1/\tau$ since $\eta(\tau)^2$ is 
transformed into $-i\tilde\tau\eta(\tilde\tau)^2$. 
The fusion rules are proposed in \cite{rational,indeco} as
\beq
\begin{array}{lll}
{\cal R}_i\times{\cal R}_j & =2{\cal R}_0+2{\cal R}_1 & i,j=0,1,\\
{\cal R}_i\times{\cal V}_j & ={\cal R}_0 & (i,j)=(0,0),(1,1),\\
& ={\cal R}_1 & (i,j)=(0,1),(1,0),\\ 
& =2{\cal V}_{-1/8}+2{\cal V}_{3/8} 
& i=0,1; j=-\frac 18,\frac 38,\\
{\cal V}_i\times{\cal V}_j & ={\cal V}_0 & (i,j)=(0,0),(1,1),\\
& ={\cal V}_1 & (i,j)=(0,1),(1,0),\\ 
& ={\cal V}_{-1/8} & (i,j)=(0,-\frac 18), (1,\frac 38),\\
& ={\cal V}_{3/8} & (i,j)=(1,-\frac 18), (0,\frac 38),\\
& ={\cal R}_0 & (i,j)=(-\frac 18,-\frac 18), (\frac 38,\frac 38),\\
& ={\cal R}_1 & (i,j)=(-\frac 18,\frac 38),(\frac 38,-\frac 18).
\end{array}
\label{eqn:fusion}
\eeq
Since the four representations ${\cal R}_0$, ${\cal R}_1$, ${\cal V}_{-1/8}$
and ${\cal V}_{3/8}$ close under the fusion, this model is considered as a 
rational conformal field theory, with a slightly weakened definition of 
rationality\cite{rational}.


\section{Boundary states of the symplectic fermion}

Now let us construct boundary states from the action (\ref{eqn:action}).
Consider a cylinder of circumference $L$ and length $T$. 
Modular parameters are defined as $\tilde q=e^{-4\pi T/L}$ and 
$\tilde\tau=2iT/L$.
As in the construction of Dirichlet and Neumann boundary states in open 
superstring theory\cite{clny,polcai,ishiono}, we assume that the boundary term 
in the action vanishes,
$
\eta\xi+\bar\eta\bar\xi=0.
$
Since $\eta$ and $\xi$ have different scaling dimensions, this condition is
decomposed into the linear conditions that 
$\eta=e^{i\phi}\bar\eta$ and $\xi=e^{i(\pi-\phi)}\bar\xi$
on the boundary.
Here, $\phi$ is a phase factor reflecting the $U(1)$ symmetry of the system.
These conditions are trivially consistent with the conformal invariance.
Although it might be possible to include non-trivial interactions between bulk 
and boundary by introducing conformally invariant boundary 
terms, we do not consider such terms here.
On Fourier decomposition of the symplectic fermion into mode operators, this
condition is rephrased as constraints on boundary states $\vert B\rangle$ 
and $\langle B\vert$
at boundaries as
\bea
&&\left(\chi^\pm_m-e^{\pm i\phi}\bar\chi^\pm_{-m}\right)\vert B\rangle=0,
\nonumber\\
&&\langle B\vert\left(\chi^\pm_m-e^{\pm i\phi}\bar\chi^\pm_{-m}\right)=0.
\label{eqn:embc}
\eea 
These equations are satisfied by coherent states,
\begin{eqnarray}
\vert B_{0\phi}\rangle 
&=& N\exp
\left(
\sum_{k>0}\frac{e^{i\phi}}{k}\chi^-_{-k}\bar\chi^+_{-k}
         +\frac{e^{-i\phi}}{k}\bar\chi^-_{-k}\chi^+_{-k}
\right)
\vert 0_\phi\rangle,\nonumber\\
\langle B_{0\phi}\vert 
&=& \langle 0_\phi \vert N^*\exp
\left(
\sum_{k>0}\frac{e^{i\phi}}{k}\chi^-_k\bar\chi^+_k
         +\frac{e^{-i\phi}}{k}\bar\chi^-_k\chi^+_k
\right),
\label{eqn:sfbs}
\end{eqnarray}
where $N$ and $N^*$ are normalization constants, and
$k$ runs over integers in the untwisted sector and half-integers in the
twisted sector. 
$\vert 0_\phi\rangle$ stands for a non-chiral ground state satisfying
\beq
\chi^\pm_{k>0}\vert 0_\phi\rangle=0=\bar\chi^\pm_{k>0}\vert 0_\phi\rangle.
\label{eqn:annih}
\eeq
In addition to this, $\vert 0_\phi\rangle$ in the untwisted sector must
satisfy
\beq
\left(\chi^\pm_0-e^{\pm i\phi}\bar\chi^\pm_0\right)\vert 0_\phi\rangle=0.
\label{eqn:em0bc}
\eeq
The bra-ground state $\langle 0_\phi\vert$ satisfies similar conditions.

The non-chiral ground state $\vert 0_\phi\rangle$ in the twisted sector is 
merely a direct product $\vert\mu\otimes\bar\mu\rangle$.
In the untwisted sector, there are two chiral bosonic ground states 
$\omega$ and $\Omega$, and therefore we have four non-chiral ground states,
namely, $\vert\omega\otimes\bar\omega\rangle$, 
$\vert\omega\otimes\bar\Omega\rangle$, $\vert\Omega\otimes\bar\omega\rangle$
and $\vert\Omega\otimes\bar\Omega\rangle$. Among these, 
$\vert\omega\otimes\bar\Omega\rangle$ and $\vert\Omega\otimes\bar\omega\rangle$
are not consistent with (\ref{eqn:em0bc}) since
$\chi^\pm_0\omega\neq 0$ and $\chi^\pm_0\Omega=0$.
The condition (\ref{eqn:em0bc}) non-trivially connects the left and right 
sectors of $\vert\omega\otimes\bar\omega\rangle$ through the 
phase factor $\phi$, whereas
$\vert\Omega\otimes\bar\Omega\rangle$ trivially satisfies (\ref{eqn:em0bc}).
Thus we have three non-chiral ground states
$\vert\omega\otimes\bar\omega\rangle$,
$\vert\Omega\otimes\bar\Omega\rangle$, and $\vert\mu\otimes\bar\mu\rangle$.
We normalize these states to be
\bea
&&\langle\omega\otimes\bar\omega\vert\omega\otimes\bar\omega\rangle=\kappa,
\nonumber\\
&&\langle\Omega\otimes\bar\Omega\vert\omega\otimes\bar\omega\rangle
=\langle\omega\otimes\bar\omega\vert\Omega\otimes\bar\Omega\rangle
=\rho\equiv -1,\nonumber\\
&&\langle\mu\otimes\bar\mu\vert\mu\otimes\bar\mu\rangle=1.
\label{eqn:gnorm}
\eea
For the convenience of later discussions we have chosen $\rho=-1$.
We leave the value of $\kappa$ unfixed.
The boundary states (\ref{eqn:sfbs}) are indexed by the phase factor $\phi$ 
and the choice of ground state $0_\phi$ which may also depend on $\phi$.
With a slight misuse of the notation we shall denote in the sequel
$\omega\otimes\bar\omega$, $\Omega\otimes\bar\Omega$ and $\mu\otimes\bar\mu$ 
as $\omega$, $\Omega$ and $\mu$, respectively. 

The condition of continuity for a general chiral field $J$ across the boundary 
$\zeta=\bar\zeta$ is
\beq
\left[J(\zeta)-\bar J(\bar\zeta)\right]_{\zeta=\bar\zeta}=0.
\label{eqn:jcont}
\eeq
A semi-annular domain in the upper half $\zeta$-plane is conformally mapped 
onto a full annulus in the $z$-plane by $z=\exp(-2\pi iw/L)$ and 
$w=(T/\pi)\ln\zeta$. Under this mapping the boundary $\zeta=\bar\zeta$ is 
mapped to
$|z|=1$, $\exp(2\pi T/L)$, and the condition (\ref{eqn:jcont}) reads
\beq
z^{s_J}J(z)=(-\bar z)^{s_J}\bar J(\bar z),
\eeq
on the boundary. Here, $s_J$ is the spin of $J$.
Now that the $z$-plane allows radial quantization, the
continuity of $J$ may be translated into conditions on the boundary states as
\cite{cardy89,ishibashi}
\beq
(J_m-(-1)^{s_J}\bar J_{-m})\vert B_{0\phi}\rangle=0.
\label{eqn:jbc}
\eeq
When $J(z)$ is the energy-momentum tensor $T(z)$,
(\ref{eqn:jbc}) reduces to the diffeomorphism invariance of the boundary
\beq
(L_m-\bar L_{-m})\vert B_{0\phi}\rangle.
\label{eqn:virbc}
\eeq  
Recalling that the Virasoro operators are given as (\ref{eqn:virop}), 
it is easily verified that indeed the states (\ref{eqn:sfbs})
satisfy this condition.
There are extra symmetries in our system other than the conformal invariance, 
as is hinted by the $U(1)$ invariance of the action (\ref{eqn:action}). 
In fact, the symmetry is upgraded to $SU(2)$ through the re-formulation 
into the symplectic fermion. Combining the $su(2)$ and the Virasoro algebra,
our system has a W-algebra of type 
${\cal W}(2,3^3)$\cite{curious,ghost,rational,hgk}, which is 
generated by the Virasoro operators (of weight 2) plus three operators of 
weight 3:
\bea
&&W^0=-\frac 12\left\{:\del\chi^+\chi^-:+:\del\chi^-\chi^+:\right\},\nonumber\\
&&W^\pm=:\del\chi^\pm\chi^\pm:,
\label{eqn:wops}
\eea
where the normal ordering is taken on appropriate (twisted or untwisted) vacuum
for each sector.
The mode operators $W^a_m$ are 
\bea
W^0_n&=&-\frac 12\sum_{j\in Z+\lambda}j
\left\{:\chi^+_{n-j}\chi^-_j:+:\chi^-_{n-j}\chi^+_j:\right\},\nonumber\\
W^\pm_n&=&\sum_{j\in Z+\lambda}j\chi^\pm_{n-j}\chi^\pm_j,
\label{eqn:wmode}
\eea
where $\lambda=0$ for the untwisted and $1/2$ for the twisted sector
as before.
Since the W-operators have spin 3, the continuity condition for the W-field
reads
\beq
(W^a_m+\bar W^a_{-m})\vert B_{0\phi}\rangle=0.
\label{eqn:wbc}
\eeq
Besides (\ref{eqn:em0bc}) which has already been
imposed for the conformal invariance, (\ref{eqn:wbc}) implies an extra
condition
\beq
e^{2i\phi}=1,
\label{eqn:z2}
\eeq
which means that the values of $\phi$ must be $0$ modulo $\pi$.
Thus in our $Z_2$ orbifold model, we restrict the values of
$\phi$ to be either $\phi=0$ or $\phi=\pi$, and in the following text we write
$\vert B_{0+}\rangle=\vert B_{0,\phi=0}\rangle$,
$\vert B_{0-}\rangle=\vert B_{0,\phi=\pi}\rangle$,
and 
$\langle B_{0+}\vert=\langle B_{0,\phi=0}\vert$,
$\langle B_{0-}\vert=\langle B_{0,\phi=\pi}\vert$. 
Then we have six distinct boundary states, 
$\vert B_{\omega+}\rangle$, $\vert B_{\omega-}\rangle$,
$\vert B_{\Omega+}\rangle$, $\vert B_{\Omega-}\rangle$,
$\vert B_{\mu+}\rangle$ and $\vert B_{\mu-}\rangle$,
which we collectively write $\vert a\rangle=\vert B_{0\phi}\rangle$.
Once boundary states are constructed, the boundary-to-boundary amplitudes
(partition functions on the cylinder) 
$\langle a\vert (\tilde q^{1/2})^{L_0+\bar L_0+1/6}\vert b\rangle$
are calculated.
Setting $|N|^2=1$, they are summarized in Table 1.

Strictly speaking, the boundary states (\ref{eqn:sfbs}) are not normalizable. 
Their non-trivial inner products are 
\bea
&&\langle B_{\omega\pm}\vert B_{\omega\mp}\rangle
=\kappa\lim_{\tilde\tau\rightarrow 0}\Lambda_{1,2}(\tilde\tau),\nonumber\\
&&\langle B_{\Omega\pm}\vert B_{\omega\mp}\rangle
=\langle B_{\omega\pm}\vert B_{\Omega\mp}\rangle=-
\lim_{\tilde\tau\rightarrow 0}\Lambda_{1,2}(\tilde\tau),\nonumber\\
&&\langle B_{\mu\pm}\vert B_{\mu\mp}\rangle
=\lim_{\tilde\tau\rightarrow 0}
(\Lambda_{0,2}(\tilde\tau)+\Lambda_{2,2}(\tilde\tau)).
\label{eqn:inprod}
\eea
The right hand sides of (\ref{eqn:inprod}) are all divergent,
which is a well-known feature of such boundary states.

\section{Modular properties of the boundary states}

The amplitude
$\langle a\vert (\tilde q^{1/2})^{L_0+\bar L_0+1/6}\vert b\rangle$
represents the tree-level graph of a closed string propagating from one 
boundary to the other. 
The same amplitude describes the one-loop graph of an open string having 
specific boundary states at each end. 
This equivalence imposes strong constraints on the {\it physical} boundary 
states. 
Assume an open string has boundary states $\tilde\alpha$ and $\tilde\beta$ at 
its ends. 
If these boundary states are physical, $n^i_{\tilde\alpha\tilde\beta}$ 
(a non-negative integer) copies 
of representations $i$ appear in the bulk. 
The partition function is then the sum of the bulk characters and written as
$Z_{\tilde\alpha\tilde\beta}(q)
=\sum_{i}n^i_{\tilde\alpha\tilde\beta}\chi_i(q)$, where $\chi_i(q)$ are the
characters for representations $i$, and $q=e^{-\pi L/T}$.
Duality of the open and closed channels demands
$Z_{\tilde\alpha\tilde\beta}(q)=\langle\tilde\alpha\vert
(\tilde q^{1/2})^{L_0+\bar L_0+1/6}\vert\tilde\beta\rangle$.
In the literature this is often called Cardy's relation\cite{cardy89}.
On expanding physical boundary states by the boundary states we constructed
above, we obtain
\beq
\sum_{i}n^i_{\tilde\alpha\tilde\beta}\chi_i(q)
=\sum_{a,b}\langle\tilde\alpha\vert a\rangle\langle a\vert
(\tilde q^{1/2})^{L_0+\bar L_0+1/6}\vert b\rangle\langle b\vert\tilde\beta\rangle.
\label{eqn:modular}
\eeq
Here, 
$\langle\tilde\alpha\vert a\rangle$ and $\langle b\vert\tilde\beta\rangle$
should be understood merely as coefficients in expansions
$\langle\tilde\alpha\vert=\sum_a\langle\tilde\alpha\vert a\rangle\langle 
a\vert$, and
$\vert\tilde\beta\rangle=\sum_b\langle b\vert\tilde\beta\rangle\vert b\rangle$,
as ordinary orthonormal bra-ket operations are not possible in our case.

Using the modular transformations of Jacobi and Dedekind 
functions\cite{jacobi}, relations between $\vert a\rangle$ and 
$\vert\tilde\alpha\rangle$ are found by comparing the coefficients of functions
on both sides. 
Some care is needed: all non-diagonal contributions have to be considered 
since on the right hand side amplitudes are not diagonalized into characters.
Equating the coefficients of $\Lambda_{1,2}(\tilde\tau)\ln\tilde q$, 
$\eta(\tilde\tau)^2\ln\tilde q$, $\Lambda_{1,2}(\tilde\tau)$, 
$\eta(\tilde\tau)^2$,
$\Lambda_{0,2}(\tilde\tau)+\Lambda_{2,2}(\tilde\tau)$ and
$\Lambda_{0,2}(\tilde\tau)-\Lambda_{2,2}(\tilde\tau)$, we have
\bea
&&\langle\tilde\alpha\vert B_{\omega+}\rangle
\langle B_{\omega-}\vert\tilde\beta\rangle
+\langle\tilde\alpha\vert B_{\omega-}\rangle
\langle B_{\omega+}\vert\tilde\beta\rangle=0,\label{eqn:cardy1}\\
&&\langle\tilde\alpha\vert B_{\omega+}\rangle
\langle B_{\omega+}\vert\tilde\beta\rangle
+\langle\tilde\alpha\vert B_{\omega-}\rangle
\langle B_{\omega-}\vert\tilde\beta\rangle
=\frac{n^{{\cal V}_0}_{\tilde\alpha\tilde\beta}
-n^{{\cal V}_1}_{\tilde\alpha\tilde\beta}}{4\pi},
\label{eqn:cardy2}\\
&&\langle\tilde\alpha\vert B_{\omega+}\rangle
\langle B_{\Omega-}\vert\tilde\beta\rangle
+\langle\tilde\alpha\vert B_{\omega-}\rangle
\langle B_{\Omega+}\vert\tilde\beta\rangle
+\langle\tilde\alpha\vert B_{\Omega+}\rangle
\langle B_{\omega-}\vert\tilde\beta\rangle
\nonumber\\
&&\;\;\;\;\;\;\;\;\;\;\;\;\;\;\;\;\;\;\;\;
+\langle\tilde\alpha\vert B_{\Omega-}\rangle
\langle B_{\omega+}\vert\tilde\beta\rangle
=n^{{\cal V}_{3/8}}_{\tilde\alpha\tilde\beta}
-n^{{\cal V}_{-1/8}}_{\tilde\alpha\tilde\beta},
\label{eqn:cardy3}\\
&&\langle\tilde\alpha\vert B_{\omega+}\rangle
\langle B_{\Omega+}\vert\tilde\beta\rangle
+\langle\tilde\alpha\vert B_{\omega-}\rangle
\langle B_{\Omega-}\vert\tilde\beta\rangle
\nonumber\\
&&\;\;\;\;
+\langle\tilde\alpha\vert B_{\Omega+}\rangle
\langle B_{\omega+}\vert\tilde\beta\rangle
+\langle\tilde\alpha\vert B_{\Omega-}\rangle
\langle B_{\omega-}\vert\tilde\beta\rangle\nonumber\\
&&\;\;\;\;
-\kappa(
\langle\tilde\alpha\vert B_{\omega-}\rangle
\langle B_{\omega-}\vert\tilde\beta\rangle
+\langle\tilde\alpha\vert B_{\omega+}\rangle
\langle B_{\omega+}\vert\tilde\beta\rangle)=0,\label{eqn:cardy4}\\
&&\langle\tilde\alpha\vert B_{\mu+}\rangle
\langle B_{\mu-}\vert\tilde\beta\rangle
+\langle\tilde\alpha\vert B_{\mu-}\rangle
\langle B_{\mu+}\vert\tilde\beta\rangle
=\frac{n^{{\cal V}_{-1/8}}_{\tilde\alpha\tilde\beta}
+n^{{\cal V}_{3/8}}_{\tilde\alpha\tilde\beta}}{2},\nonumber\\
&&\label{eqn:cardy5}\\
&&\langle\tilde\alpha\vert B_{\mu+}\rangle
\langle B_{\mu+}\vert\tilde\beta\rangle
+\langle\tilde\alpha\vert B_{\mu-}\rangle
\langle B_{\mu-}\vert\tilde\beta\rangle
\nonumber\\
&&\;\;\;\;\;\;\;\;\;\;\;\;\;\;\;\;\;\;\;\;\;\;\;\;\;\;\;\;
=n^{{\cal R}_0}_{\tilde\alpha\tilde\beta}
+n^{{\cal R}_1}_{\tilde\alpha\tilde\beta}
+\frac 14 (n^{{\cal V}_0}_{\tilde\alpha\tilde\beta}
+n^{{\cal V}_1}_{\tilde\alpha\tilde\beta}).\label{eqn:cardy6}
\eea
Now following Cardy's fusion method\cite{cardy89}, 
let us find the consistent physical boundary states one by one.
We assume that bra and ket boundary states have the same real
coefficients
$\langle a\vert\tilde\alpha\rangle=\langle\tilde\alpha\vert a\rangle$.
We start by looking for a reference state $\vert\tilde{\cal V}_0\rangle$ 
such that $n^i_{\tilde{\cal V}_0\tilde\alpha}=
n^i_{\tilde\alpha\tilde{\cal V}_0}=\delta^i_{\tilde\alpha}$.
Let $\tilde\alpha=\tilde\beta=\tilde{\cal V}_0$ in 
(\ref{eqn:cardy1})-(\ref{eqn:cardy6}). 
In the untwisted sector, from (\ref{eqn:cardy1}) we have
$\langle\tilde{\cal V}_0\vert B_{\omega+}\rangle
\langle B_{\omega-}\vert\tilde{\cal V}_0\rangle=0$. Since we can exchange
$\phi=0$ and $\phi=\pi$ as a consequence of $Z_2$ symmetry,
we may put 
$\langle\tilde{\cal V}_0\vert B_{\omega+}\rangle
=\langle B_{\omega+}\vert\tilde{\cal V}_0\rangle
=0$ without loss of generality. 
Then from (\ref{eqn:cardy2}) we have 
$\vert\langle\tilde{\cal V}_0\vert B_{\omega-}\rangle\vert^2=1/(4\pi)$, so
$\langle\tilde{\cal V}_0\vert B_{\omega-}\rangle
=\langle B_{\omega-}\vert\tilde{\cal V}_0\rangle=1/(2\sqrt\pi)$.
Substituting these values, (\ref{eqn:cardy3}) gives 
$\langle\tilde{\cal V}_0\vert B_{\Omega+}\rangle
=\langle B_{\Omega+}\vert\tilde{\cal V}_0\rangle=0$.
From (\ref{eqn:cardy4}) we find
$\langle\tilde{\cal V}_0\vert B_{\Omega-}\rangle
=\langle B_{\Omega-}\vert\tilde{\cal V}_0\rangle=\kappa/(4\sqrt\pi)$.
In the twisted sector, (\ref{eqn:cardy5}) becomes 
$\langle\tilde{\cal V}_0\vert B_{\mu+}\rangle
\langle B_{\mu-}\vert\tilde{\cal V}_0\rangle=0$, and again without losing
generality we can choose 
$\langle\tilde{\cal V}_0\vert B_{\mu+}\rangle
=\langle B_{\mu+}\vert\tilde{\cal V}_0\rangle=0$.
Then from (\ref{eqn:cardy6}) we find
$\langle\tilde{\cal V}_0\vert B_{\mu-}\rangle
=\langle B_{\mu-}\vert\tilde{\cal V}_0\rangle=1/2$.
Thus we found 
$\vert\tilde{\cal V}_0\rangle
=(1/2\sqrt\pi)\vert B_{\omega-}\rangle
+(\kappa/4\sqrt\pi)\vert B_{\Omega-}\rangle
+(1/2)\vert B_{\mu-}\rangle$.

Next, we put $\tilde\alpha=\tilde{\cal V}_1$ and 
$\tilde\beta=\tilde{\cal V}_0$.
We find 
$\langle\tilde{\cal V}_1\vert B_{\omega-}\rangle
=\langle B_{\omega-}\vert\tilde{\cal V}_1\rangle=-1/(2\sqrt\pi)$,
$\langle\tilde{\cal V}_1\vert B_{\Omega-}\rangle
=\langle B_{\Omega-}\vert\tilde{\cal V}_1\rangle=-\kappa/(4\sqrt\pi)$,
$\langle\tilde{\cal V}_1\vert B_{\mu-}\rangle
=\langle B_{\mu-}\vert\tilde{\cal V}_1\rangle=1/2$,
and
$\langle\tilde{\cal V}_1\vert B_{\omega+}\rangle
=\langle B_{\omega+}\vert\tilde{\cal V}_1\rangle
=\langle\tilde{\cal V}_1\vert B_{\Omega+}\rangle
=\langle B_{\Omega+}\vert\tilde{\cal V}_1\rangle
=\langle\tilde{\cal V}_1\vert B_{\mu+}\rangle
=\langle B_{\mu+}\vert\tilde{\cal V}_1\rangle=0$.
The rest of the states are found similarly by putting 
$\tilde\alpha=\tilde{\cal V}_{-1/8},\;\tilde{\cal V}_{3/8},\;
\tilde{\cal R}_0,\;\tilde{\cal R}_1$ one by one, all with 
$\tilde\beta=\tilde{\cal V}_0$.
Then we find 
\bea
&&\vert\tilde{\cal V}_0\rangle
=\frac{1}{2\sqrt{\pi}}\vert B_{\omega-}\rangle
+\frac{\kappa}{4\sqrt{\pi}}\vert B_{\Omega-}\rangle
+\frac 12 \vert B_{\mu-}\rangle,\nonumber\\
&&\vert\tilde{\cal V}_1\rangle
=\frac{-1}{2\sqrt{\pi}}\vert B_{\omega-}\rangle
-\frac{\kappa}{4\sqrt{\pi}}\vert B_{\Omega-}\rangle
+\frac 12 \vert B_{\mu-}\rangle,\nonumber\\
&&\vert\tilde{\cal V}_{-1/8}\rangle
=\vert B_{\mu+}\rangle
-2\sqrt{\pi}\vert B_{\Omega+}\rangle,\nonumber\\
&&\vert\tilde{\cal V}_{3/8}\rangle
=\vert B_{\mu+}\rangle
+2\sqrt{\pi}\vert B_{\Omega+}\rangle,\nonumber\\
&&\vert\tilde{\cal R}\rangle
\equiv\vert\tilde{\cal R}_0\rangle
=\vert\tilde{\cal R}_1\rangle
=2\vert B_{\mu-}\rangle.
\label{eqn:pbsket}
\eea
Since $\tilde{\cal R}_0$ and $\tilde{\cal R}_1$ are the same state,
we shall denote it as $\tilde{\cal R}$.
There are other solutions obtained from 
the above by exchanging 
$\omega+$ and $\omega-$, $\Omega+$ and $\Omega-$, $\mu+$ and $\mu-$
(first and second pairs have to be exchanged simultaneously),
as a consequence of the $Z_2$ symmetry.
Apart from this, the solutions are unique. 
Therefore, the duality of open and closed string channels provides strong 
enough constraints for the physical boundary states to be determined without
ambiguity. Substituting these states back into (\ref{eqn:modular}), 
possible $n^i_{\tilde\alpha\tilde\beta}$ on the left hand side are found.
Note that $n^i_{\tilde\alpha\tilde\beta}$ cannot be determined uniquely by this
procedure, since the characters are not independent but
$\chi_{{\cal R}_0}=\chi_{{\cal R}_1}=2(\chi_{{\cal V}_0}+\chi_{{\cal V}_1})$.
Up to this ambiguity $n^i_{\tilde\alpha\tilde\beta}$ is identical to the 
fusion matrix $n^i_{jk}$, 
which appears as $\Phi_j\times\Phi_k=\sum_i n^i_{jk}\Phi_i$
($\Phi$'s stand for the representations) in the fusion rule (\ref{eqn:fusion}).

\section{Discussion}

The essential point in our analysis is the appearance of the term 
$\eta(\tilde\tau)^2\ln\tilde q$ in the cylinder amplitude (Table 1)
through the proper treatment of the zero-mode.
Note that the {\em five} modular functions $\eta(\tilde\tau)^2$,
$\eta(\tilde\tau)^2\ln\tilde q$, $\Lambda_{0,2}(\tilde\tau)$,
$\Lambda_{1,2}(\tilde\tau)$, $\Lambda_{2,2}(\tilde\tau)$ close under the 
modular transformation $\tilde\tau\rightarrow -1/\tilde\tau$.
Discarding either $B_{\omega+}$ or $B_{\omega-}$ in order to get rid of
the unwanted function $\Lambda_{1,2}(\tilde\tau)^2\ln\tilde q$, 
we obtained a set of boundary states including the reference state 
$\tilde{\cal V}_0$ which is necessary for the Cardy fusion procedure.
This situation is quite similar to what happens in the Ising
model case\cite{leclair,yamaguchi,nepo}, 
where one of the two $R$ sector states has to be discarded to
give three boundary states, namely spin up, down, and free, which behave
appropriately under modular transformations. 
In fact, the Ising model is the only known example which allows free-field 
construction of boundary states.

However, our model differs from the Ising model in one important respect.
Neglecting the row and column involving the discarded state $B_{\omega+}$, 
the cylinder amplitude of the untwisted sector in Table 1 gives a matrix
\beq
\left(
\begin{array}{ccc}
(\kappa-\ln\tilde q)\eta(\tilde\tau)^2&-\Lambda_{1,2}(\tilde\tau)
&-\eta(\tilde\tau)^2\\
-\Lambda_{1,2}(\tilde\tau)&0&0\\
-\eta(\tilde\tau)^2&0&0
\end{array}
\right),
\eeq
which is not regular. Since one of the three eigen-values is zero, the
untwisted sector has only two non-trivial partition functions on 
diagonalization. This means that the net content of the space spanned by
$\vert B_{\omega-}\rangle$, $\vert B_{\Omega\pm}\rangle$, 
$\vert B_{\mu\pm}\rangle$
consists of only four states, not five. 
Therefore it is {\it not} possible to allocate five boundary states to the 
five modular functions.

This is related to the difficulty in expressing the physical boundary states
in terms of the Ishibashi states.
In ordinary CFTs, the solutions to (\ref{eqn:virbc}) are found in the 
form of Ishibashi states\cite{ishibashi}
\beq
\vert j\rishi\equiv\sum_M\vert j;M\rangle\otimes U\overline{\vert j;M\rangle},
\label{eqn:ishibashi}
\eeq
where $U$ is an antiunitary operator, $j$ is a label for Verma modules, 
and $M$ is a level in the module. Ishibashi states diagonalize the cylinder amplitudes to 
give characters, 
\bea
&&\lishi i\vert(\tilde q^{1/2})^{L_0+\bar L_0-c/12}\vert i\rishi
=\chi_i(\tilde q),
\label{eqn:ishidef}\\
&&\lishi i\vert(\tilde q^{1/2})^{L_0+\bar L_0-c/12}\vert j\rishi=0
\;\;\mbox{for}\;\; i\neq j.
\label{eqn:ishiortho}
\eea
In our model, we can find candidates for the Ishibashi states such as
\bea
&&\vert{\cal V}_0\rishi
=\frac 12\vert B_{\Omega+}\rangle
+\frac 12\vert B_{\Omega-}\rangle\nonumber,\\
&&\vert{\cal V}_1\rishi
=\frac 12\vert B_{\Omega+}\rangle
-\frac 12\vert B_{\Omega-}\rangle\nonumber,\\
&&\vert{\cal V}_{-1/8}\rishi
=\frac 12\vert B_{\mu+}\rangle
+\frac 12\vert B_{\mu-}\rangle\nonumber,\\
&&\vert{\cal V}_{3/8}\rishi
=\frac 12\vert B_{\mu+}\rangle
-\frac 12\vert B_{\mu-}\rangle,\nonumber\\
&&\vert{\cal R}\rishi
\equiv\vert{\cal R}_0\rishi
=\vert{\cal R}_1\rishi
=\sqrt{2}\vert B_{\Omega+}\rangle,
\label{eqn:ishiket}
\eea
and
\bea
&&\lishi{\cal V}_0\vert
=\frac{-1}{2}\langle B_{\omega-}\vert
-\frac 12\langle B_{\omega+}\vert\nonumber,\\
&&\lishi{\cal V}_1\vert
=\frac 12\langle B_{\omega+}\vert
-\frac 12\langle B_{\omega-}\vert\nonumber,\\
&&\lishi{\cal V}_{-1/8}\vert
=\frac 12\langle B_{\mu+}\vert
+\frac 12\langle B_{\mu-}\vert\nonumber,\\
&&\lishi{\cal V}_{3/8}\vert
=\frac 12\langle B_{\mu-}\vert
-\frac 12\langle B_{\mu+}\vert\nonumber,\\
&&\lishi{\cal R}\vert
\equiv\lishi{\cal R}_0\vert
=\lishi{\cal R}_1\vert
=-\sqrt{2}\langle B_{\omega-}\vert,
\label{eqn:ishibra}
\eea
whereby the characters (\ref{eqn:char}) are reproduced in the form 
(\ref{eqn:ishidef}), and the orthogonality (\ref{eqn:ishiortho}) holds for
${\cal V}_0$, ${\cal V}_1$, ${\cal V}_{-1/8}$, and ${\cal V}_{3/8}$. 
Note that it is not possible to find such states with the 
same bra and ket coefficients.
It can be easily checked that the physical boundary states $\tilde{\cal V}_0$
and $\tilde{\cal V}_1$ cannot be expressed as linear combinations of the 
states (\ref{eqn:ishiket}), (\ref{eqn:ishibra}). 
As a consequence, it is not possible to derive the Verlinde formula using
the modular transformations as in \cite{cardy89}, since $\tilde{\cal V}_0$
plays an essential role in such discussions. The failure of the Verlinde
formula is indeed consistent with the fusion rule (\ref{eqn:fusion}),
which cannot be diagonalised.

Alternatively, the {\it four} representations ${\cal R}_0$, ${\cal R}_1$, 
${\cal V}_{-1/8}$ and ${\cal V}_{3/8}$ can be regarded as fundamental
constituents of the theory, since they themselves close under the fusion.
It is argued by Kausch and Gaberdiel\cite{local} that local and non-chiral 
bulk theory with finite multiplicity is given by three non-chiral 
representations, namely, ${\cal V}_{-1/8}\otimes\bar{\cal V}_{-1/8}$,
${\cal V}_{3/8}\otimes\bar{\cal V}_{3/8}$, and
${\cal R}$ which is a combination of 
$({\cal R}_0\otimes\bar{\cal R}_0)/{\cal N}_{0\bar 0}$ and
$({\cal R}_1\otimes\bar{\cal R}_1)/{\cal N}_{1\bar 1}$, where 
${\cal N}_{0\bar 0}$ and ${\cal N}_{1\bar 1}$ are subspaces to be
quotiented out.
This is analogous to our result that the physical boundary 
states for ${\cal R}_0$ and ${\cal R}_1$ are identical.
Considering the four representations ${\cal R}_0$, ${\cal R}_1$, 
${\cal V}_{-1/8}$ and ${\cal V}_{3/8}$, 
we see from (\ref{eqn:pbsket}) and (\ref{eqn:ishiket}) that
the physical ket-states and Ishibashi ket-states are related as
\bea
&&\vert\tilde{\cal R}\rangle
=2\vert{\cal V}_{-1/8}\rishi-2\vert{\cal V}_{3/8}\rishi,\nonumber\\
&&\vert\tilde{\cal V}_{-1/8}\rangle
=\vert{\cal V}_{-1/8}\rishi+\vert{\cal V}_{3/8}\rishi
-\sqrt{2\pi}\vert{\cal R}\rishi,\nonumber\\
&&\vert\tilde{\cal V}_{3/8}\rangle
=\vert{\cal V}_{-1/8}\rishi+\vert{\cal V}_{3/8}\rishi
+\sqrt{2\pi}\vert{\cal R}\rishi.
\label{eqn:brarel}
\eea  
These are the combinations of $\vert B_{\mu\pm}\rangle$ and 
$\vert B_{\Omega+}\rangle$.
However, the boundary bra-states for these representations cannot be expressed 
in terms of the corresponding Ishibashi bra-states (\ref{eqn:ishibra}),
since the former are the combinations of $\langle B_{\mu\pm}\vert$ and
$\langle B_{\Omega+}\vert$, whereas the latter are of 
$\langle B_{\mu\pm}\vert$ and $\langle B_{\omega-}\vert$.
The candidate of the Ishibashi states (\ref{eqn:ishiket}), (\ref{eqn:ishibra})
are not unique, and alternatively, we can define such states so that the 
bra-states are linearly related to the physical boundary states,
but then the ket-states cannot be.
That is, it is possible to express the physical boundary states in terms of 
such Ishibashi states on either of the two boundaries, but not on both.

We started from the Lagrangian representation of the $c=-2$ LCFT model
and presented a possible solution for physical boundary states. 
Modular invariance imposes tight enough constraints on the partition 
function to identify the boundary states which allow the appearance of bulk 
representations.
Although we could find five possible physically consistent boundary states
$\tilde{\cal R}$,
$\tilde{\cal V}_0$, $\tilde{\cal V}_1$,
$\tilde{\cal V}_{-1/8}$ and $\tilde{\cal V}_{3/8}$,
their implication is still not evident.
The three states $\tilde{\cal R}$, $\tilde{\cal V}_{-1/8}$, 
$\tilde{\cal V}_{3/8}$ may be considered as
{\it genuinely} physical as they correspond to non-chiral bulk 
representations.
However, this speculation is not necessarily persuasive.
Among well-studied unitary minimal models, the 3-state Potts model is known to
possess a W-algebra, and its complete boundary states were found quite 
recently\cite{3sp-aos,3sp-fs,3sp-bpz}.
In that model, only the fixed and mixed boundary states are 
obtained by Cardy's method from the W-invariant conformal towers; in order 
to obtain the complete set including ``free'' and ``new'' boundary states, 
all chiral representations from the Kac table not constrained by the 
W-symmetry had to be considered. 
Then in our $c=-2$ model, the boundary states $\tilde{\cal V}_0$ and 
$\tilde{\cal V}_1$ may well represent some boundary conditions of 
a statistical model since the $c_{2,1}$ Kac table indicates representations 
with conformal weights $0$, $1$, $-1/8$ and $3/8$. 
Further study of the properties of the states we have obtained
requires analysis of particular statistical models, such as critical 
polymers\cite{saleur,dupsal}. These topics are beyond the scope of this paper.

\acknowledgments

The authors appreciate stimulating discussions with I. I. Kogan and
Y. Ishimoto.



\onecolumn
{\samepage
\begin{center}
Table 1: Amplitudes 
$\langle a\vert (\tilde q^{1/2})^{L_0+\bar L_0+1/6}\vert b\rangle$.
\end{center}
\begin{center}
\underline{Untwisted Sector}
\\
\begin{tabular}{c|c|c|c|c} 
\hline
&\multicolumn{4}{c}{$\vert b\rangle$}\\
\cline{2-5}
$\langle a\vert$&$B_{\omega+}$&$B_{\omega-}$&$B_{\Omega+}$&$B_{\Omega-}$\\
\hline\hline
$B_{\omega+}$&$(\kappa-\ln\tilde q)\eta(\tilde\tau)^2$&
$\;\;\;(\kappa-\ln\tilde q)\Lambda_{1,2}(\tilde\tau)\;\;\;$&
$-\eta(\tilde\tau)^2$&
$\;\;\;-\Lambda_{1,2}(\tilde\tau)\;\;\;$\\
\hline
$B_{\omega-}$&
$\;\;\;(\kappa-\ln\tilde q)\Lambda_{1,2}(\tilde\tau)\;\;\;$&
$(\kappa-\ln\tilde q)\eta(\tilde\tau)^2$&
$\;\;\;-\Lambda_{1,2}(\tilde\tau)\;\;\;$&
$-\eta(\tilde\tau)^2$\\
\hline
$B_{\Omega+}$&
$-\eta(\tilde\tau)^2$&
$-\Lambda_{1,2}(\tilde\tau)$&
$0$&$0$\\
\hline
$B_{\Omega-}$&
$-\Lambda_{1,2}(\tilde\tau)$&
$-\eta(\tilde\tau)^2$&$0$&$0$\\
\hline
\end{tabular}
\end{center}

\begin{center}
\underline{Twisted Sector}
\\
\begin{tabular}{c|c|c} 
\hline
&\multicolumn{2}{c}{$\vert b\rangle$}\\
\cline{2-3}
$\langle a\vert$&$B_{\mu+}$&$B_{\mu-}$\\
\hline\hline
$B_{\mu+}$&$\;\;\;\Lambda_{0,2}(\tilde\tau)-\Lambda_{2,2}(\tilde\tau)\;\;\;$&
$\;\;\;\Lambda_{0,2}(\tilde\tau)+\Lambda_{2,2}(\tilde\tau)\;\;\;$\\
\hline
$B_{\mu-}$&$\Lambda_{0,2}(\tilde\tau)+\Lambda_{2,2}(\tilde\tau)$&
$\Lambda_{0,2}(\tilde\tau)-\Lambda_{2,2}(\tilde\tau)$\\
\hline
\end{tabular}
\end{center}}
\end{document}